\documentclass{ws-procs9x6}
\usepackage{amssymb}

\begin{document}

\title{Understanding High $T_c$ Superconductivity\footnote{\uppercase{I}nvited talk at the \uppercase{I}nternational 
\uppercase{S}ymposium on \uppercase{F}rontiers of \uppercase{S}cience in celebration of the 80th birthday of \uppercase{P}rof. 
\uppercase{C}hen \uppercase{N}ing \uppercase{Y}ang}.}
\author{Zheng-Yu Weng}
\address{Center for Advanced Study, Tsinghua University\\
Beijing, 100084, China\\ 
E-mail: weng@castu.tsinghua.edu.cn}
\maketitle


\abstracts{I briefly review several key issues in understanding the cuprate superconductors
from the point of view of doped-Mott-insulator. Then I present an effective
low-energy theory and show that the phase diagram
of such a model includes an antiferromagnetic (AF) phase, a superconducting
state, and a pseudogap phase at low doping and low temperature, consistent
with the experimental one. The dual topological gauge structure of this
model is responsible for unifying these phases within a single framework
with correct doping and temperature scales. Some unique predictions in such
a model are also discussed.}

\section{Introduction}

High-temperature superconductivity has been discovered in the cuprate
materials for more than a decade by now. These materials are widely
considered to be doped Mott insulators \cite{pwa}. They become Mott
insulators in the so-called half-filling case where each lattice site in
copper-oxide layers is occupied by and only by one conduction electron. The
superconducting phase emerges in the doped case as the half-filled electrons
in the copper-oxide layers are pumped out (or equivalently, `holes' are
injected into the system). The normal state above the superconducting
transition temperature also exhibits anomalous phenomena distinctively
different from a conventional metal under a Landau-Fermi liquid description.

How to understand the mechanism of superconductivity and other novel
properties, which have been experimentally found in almost all probing
channels, including the single-particle, magnetic, transport, and pairing
channels, etc., has posed a great challenge for the condensed matter
physicists. In spite of tremendous experimental and theoretical efforts,
with a lot of theoretical proposals, there still lacks a consensus by far on
what is the correct theory or sensible theoretical framework. Nevertheless,
a great number of researchers in this field now believe that such materials
belong to strongly correlated electron systems and the fundamental physics
of a doped Mott insulator may hold the key to the heart of main issues in
high-temperature superconductivity.

\emph{Doped Mott insulators. }A Mott insulator is purely an interaction
effect as the strong on-site Coulomb repulsion prevents two electrons from
staying at the same site, such that the charge degrees of freedom are
essentially frozen. In the doped case, as the electron number is reduced to
less than one on average at each site, the charge degrees of freedom are no
longer completely frozen and electrons can hop to empty sites without a
penalty from the Coulomb potential. However, the majority of the charge
degrees of freedom of the electrons still remains frozen at small doping and
in general electrons cannot be simply viewed as quasiparticles with a long
life time near the Fermi surface. In particular, a quasiparticle in
conventional Fermi liquids carries both charge and spin. But in the
half-filling limit of the Mott insulator, the remaining low-lying degrees of
freedom of the electrons only involve spins. Such a separation of two
degrees of freedom is also obvious at small doping, where the charge carrier
number $x$ is much less than the number of spins, $1-x,$ per site (which is
equal to the electron number). The spin-charge separation concept \cite
{pwa1,KRS,zou} is therefore the most essential basis in understanding the
Mott physics.

\emph{Spin-charge separation description. }But an oversimplified spin-charge
separation picture, namely, the charge carriers (`holons') and neutral spin
carriers (`spinons') are completely decoupled, are not consistent with
experiments as well as theoretical considerations, although the overall
counting of charge and spin numbers of the doped Mott insulator are
correctly incorporated. It turns out that there still exists a strong
coupling between holons and spinons, reflecting the fact that the Hilbert
space is restricted due to the on-site Coulomb interaction. Generally such a
type of interactions is present in a form of gauge fields (This shows how
interestingly a gauge theory can emerge in condensed matter systems with
strong correlations of electrons) \cite{gauge}.

The spin-charge separation and the gauge interactions constitute a very
basic framework for describing a doped Mott insulator. This has been
generally accepted. But it is not yet the whole story. One still needs to
identify the (short-range) spin correlations, which will provide a driving
force behind the low-lying charge and spin dynamics. This is where various
theories of doped Mott insulators differ from each other, in essence.

\emph{RVB states. }At half-filling, the low-lying spin degrees of freedom of
the cuprates are well described by the Heisenberg model, given by

\begin{equation}
H_J=J\sum_{\left\langle ij\right\rangle }\mathbf{S}_i\cdot \mathbf{S}_j+%
\text{const.}  \label{hj}
\end{equation}
Here $\left\langle ij\right\rangle $ denotes the nearest-neighbor (NN) sites.
This model predicts a long-range N\'{e}el order in the ground state and
spin-wave excitations in long-wavelength, low-energy regime, consistent with
the experiment. However, for the purpose of studying the doped case, strong
quantum correlations at \emph{short-range} are actually more important. This
is because the hole hopping is usually quite sensitive to the short-range
spin correlations in the background. For example, in the most widely used
model of doped Mott insulator, \emph{i.e}., the $t-J$ model, besides the
superexchange term $H_J,$ the hopping term is given by

\begin{equation}
H_t=-t\sum_{\left\langle ij\right\rangle }c_{i\sigma }^{\dagger }c_{j\sigma
}+h.c.  \label{ht}
\end{equation}
under the no double occupancy constraint 
\begin{equation}
\sum_ic_{i\sigma }^{\dagger }c_{i\sigma }\leq 1.  \label{constraint}
\end{equation}
Here the hole hopping involves the NN sites, which is strongly influenced by
the short-range spin correlations. To correctly describe the doping effect,
therefore, a correct description of short-range spin correlations is
essential.

Different kinds of short-range spin correlations can lead to drastically
different low-energy physics. These short-range spin correlations will then
set a natural `ultraviolet' (high-energy) cutoff for a low-energy
effective model of doped Mott insulators, at least in lightly doped
regime, and provide the driving force for the system's low-energy behavior.
In contrast, the \emph{long-range} AF correlations of spins will quickly and
easily get suppressed as a small amount of holes are added to the system,
and will usually play only a less important and respondent role in the doped
regime.

Anderson proposed a resonating-valence-bond (RVB) picture to describe such
short-range quantum correlations of spins, in which spins form singlet pairs
and superpose coherently to form an RVB background. Once the RVB state is
given, the low-energy effective model for doped Mott insulators is usually
fixed through a spin-charge separation and gauge-theory formulation.
Therefore, how to correctly describe the short-range spin correlations is
the most essential issue in constructing a sensible theory for the cuprates
based on the doped Mott insulator scenario.

So far there are roughly two kinds of RVB states being proposed. In the
first kind, spins (spinons) are fermions and they form RVB pairs \cite
{pwa,baskaran}. The theoretical structure is quite similar to the BCS
theory, except that the charged Cooper pairs are replaced by neutral spinon
pairs. The mathematical advantage of this description is the familiarity to
the BCS theory. But such a state notably fails to describe the correct
long-range AF correlations at half-filling. Nevertheless, people have hoped
that it may capture the right physics of short-ranged correlations and thus
could become more relevant in the doped case when the AF long-range order is
gone and superconductivity sets in. However, as we shall argue below, this
may well not be the case even at finite doping.

The second kind is known as the bosonic RVB state. Here spins (spinons) are
bosons. A variational wavefunction based on the bosonic RVB picture can
produce\cite{lda} an unrivaled accurate ground-state energy ($-0.3344J$ per
bond as compared to the exact numerical value of $-0.3346J$ per bond for the
Heisenberg model), and a generalized calculation\cite{chen} can precisely
provide not only the ground-state energy, staggered magnetization, but also
spin excitation spectrum in the whole Brillouin zone. Since the energy of
the Heisenberg model is directly related to the NN spin-spin correlations, a
good variational energy also means a good description of short-range spin
correlations. By contrast, a variational energy in the fermionic RVB state
gives rise to a number, $-0.319J$ per bond, which suggests that the
description of the short-range spin correlations is less accurate. In fact,
in the short-range spin-spin correlations of the fermionic RVB state, the AF
component is even worse (one finds that the majority contribution to the
energy is not from the AF correlations as opposed to the exact numerics and
the bosonic RVB theory).

\emph{Doping: the phase string effect. }Due to the absence of the
short-range AF correlations, the hopping of holes is not very frustrated in
the fermionic RVB description. In contrast, in the bosonic RVB description,
the coherent motion of doped holes is much more difficult. For example, it
can easily show\cite{1-hole} that the hopping integral vanishes on average in the
half-filling bosonic RVB state, and thus the kinetic energy is highly
frustrated for a hole to move on such an AF background. This kinetic energy
frustration reflects the sharp difference between the fermionic and bosonic
RVB states in the doped case.

The underlying physics can be traced back to a fundamental property of the $%
t-J$ model, known as the phase string effect \cite{string1}. It has been
well known that the ground state wavefunction of the Heisenberg model
satisfies the Marshall sign \cite{marshall}. Such a Marshall sign rule would
hold even at arbitrary doping, if holes remain static on lattice sites. But
once the holes start to move, the Marshall sign rule will be scrambled by
the hopping of a hole on its path. Such a disordered Marshall sign can be
described by a product of sequential $+$ and $-$ signs, $(+1)\times
(-1)\times (-1)\times ...$, where signs are determined by simply counting
the index $\sigma $ of each spin exchanged with the hole during its hopping%
\cite{string1}. Physically a phase string can be regarded as the transverse
mismatches of spins created by the motion of the hole. The significance of
such a phase string is that it represents the \textit{sole} source of phase
frustrations in the $t-J$ model, which cannot be `repaired' at low energy,
and therefore is the most essential doping effect in such a doped Mott
insulator.

\emph{Mutual gauge interactions. }The phase string effect indicates an
intrinsic interaction between the spin and charge degrees of freedom. At
small doping limit, without changing the spin background, which remains to
be an antiferromagnet, the doped holes will pick up the phase string effect
which strongly frustrates their kinetic energy. Here the total sign $%
(+1)\times (-1)\times (-1)\times ...$ will become uncertain with the
increase of the length of paths, leading to the localization of the holes
(see below)\cite{1-hole}. At larger doping, the spin background has to be eventually
re-adjusted to minimize the frustration of the phase string effect in favor
of the kinetic energy of holes. Note that the phase string $(+1)\times
(-1)\times (-1)\times ...$ depends on those spins exchanged with the hole
during its hopping. So if the background spins are tightly in RVB pairing,
the randomness in a phase string will be qualitatively removed and holes
will then become delocalized. The corresponding ground state will be shown
to be superconducting. On the other hand, the long-range AF order generally
has to disappear in such a state.

Therefore, in the following we shall take the point of view that the cuprate
superconductors are essentially doped Mott insulators. In order to describe
a doped Mott insulator, we introduce a spin-charge separation
framework with an intrinsic gauge structure. We adopt the bosonic RVB
description to characterize the short-range, high-energy spin correlations,
which works extremely well at half-filling and provides an essential
underpinning for the low-energy theory at both half-filling and small
doping. We identify a nonlocal mutual interaction between the spin and
charge degrees of freedom, {\em i.e.}, the phase string effect, which will
result in a topological gauge structure in the spin-charge separation
formulation of the whole problem.

In the remaining part of the talk, I shall first present the low-energy
effective model established based on the above-outlined considerations. We
call it the phase string theory. I shall then present a series of important
results predicted by this model. I shall show that this theory naturally
unifies antiferromagnet and superconductivity within a single theoretical
framework and its unique topological structure determines a rich, novel
phase diagram in a self-consistent fashion, which is also in good agreement
with the experimental measurements of the cuprate superconductors.

\section{Phase string theory}

\emph{Phase string model. }The low-energy effective Hamiltonian with
incorporating the bosonic RVB pairing and the phase string effect induced by
doping can be derived \cite{string2} from the $t-J$ model as follows: 
\[
H_{string}=H_h+H_s
\]
with 
\begin{eqnarray}
H_h &=&-t_h\sum_{\langle ij\rangle }(e^{iA_{ij}^s})h_i^{\dagger }h_j+H.c.  \label{hh} \\
H_s &=&-J_s\sum_{\langle ij\rangle \sigma }(e^{i\sigma A_{ij}^h})b_{i\sigma
}^{\dagger }b_{j-\sigma }^{\dagger }+H.c.  \label{hs}
\end{eqnarray}
Here $h_i$ and $b_{i\sigma }$ are \emph{bosonic} `holon' and `spinon'
operators, respectively. An electron operator is composed of such spinless
holon and neutral spinon operators by $c_{i\sigma }=h_i^{\dagger }b_{i\sigma
}e^{i\hat{\Theta}_{i\sigma }}.$ Note that the fermionic commutation
relations of $c_{i\sigma }$'s are guaranteed by the phase factor $e^{i\hat{%
\Theta}_{i\sigma }}$, which is defined by $e^{i\hat{\Theta}_{i\sigma
}}=(-\sigma )^ie^{i\frac 12\left[ \Phi _i^b-\sigma \Phi _i^h\right] }~,$
with $\Phi _i^b=\Phi _i^s-\Phi _i^0,~$where $\Phi _i^s=\sum_{l\neq i}\theta
_i(l)\sum\nolimits_\alpha \alpha n_{l\alpha }^b,$ $\Phi _i^0=\sum_{l\neq
i}\theta _i(l),~$and $\Phi _i^h=\sum_{l\neq i}\theta _i(l)n_l^h~$($%
n_{l\alpha }^b$ and $n_l^h$ are spinon and holon number operators,
respectively, and $\theta _i(l)\equiv \mbox{Im ln $(z_i-z_l)$}$, where $%
z_l=x_l+iy_l$ is the complex coordinate of a lattice site $l$).

At low doping, $t_h\sim t$ and $J_s=1/2\Delta ^sJ\sim J$ ($t$ and $J$ are
the parameters in the $t-J$ model). Here the effective theory is underpinned
by the bosonic RVB order parameter $\Delta ^s,$ which characterizes the
short range (NN) AF correlations as $\langle \mathbf{S}_i\cdot \mathbf{S}%
_j\rangle _{NN}=-3/8|\Delta ^s|^2$. In the following we shall denote the
transition temperature for $\Delta ^s\neq 0$ by $T_0$. At small doping, $%
\Delta ^s\neq 0$ covers a temperature regime extended over $T_0\sim 1,000$K.
With the increase of doping, $T_0$ is expected to monotonically decrease as
illustrated in Fig. 1, showing an applicable region for the present
effective theory. Beyond this region, the short-range AF correlations
disappear and the novel properties described by (\ref{hh}) and (\ref{hs}),
including the superconductivity, are no longer present.

\begin{figure}[t]
\centerline{\epsfxsize=3.9in\epsfbox{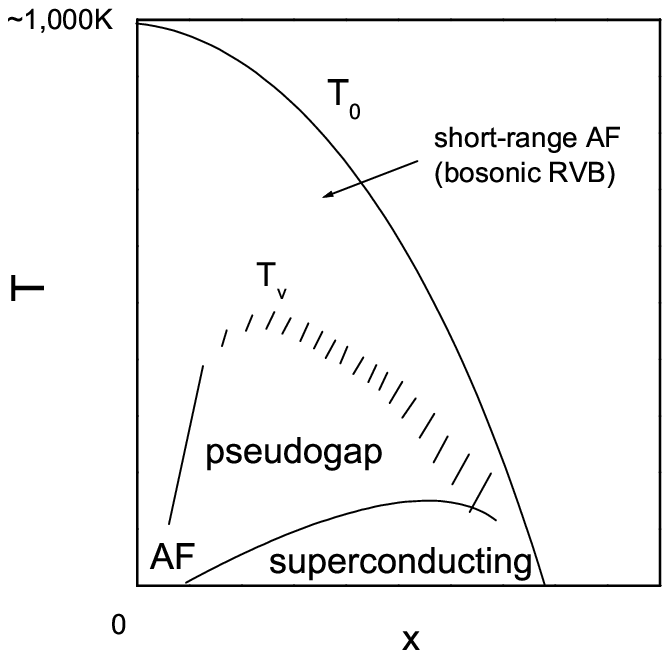}}
\caption{The phase string model is underpinned by the bosonic RVB pairing 
($\Delta^s\neq 0$), characterizing short-range AF correlations, in the regime below $T_0$, 
which covers a series of phases including AF, pseudogap, and superconducting states. }
\label{inter}
\end{figure}

\emph{Topological gauge structure.} $H_{string}$ is invariant under the
gauge transformation: $h_j\rightarrow h_j\exp (i\varphi _j),$ $%
A_{ij}^s\rightarrow A_{ij}^s+i(\varphi _i-\varphi _j)$ and $b_{j\sigma
}\rightarrow b_{j\sigma }\exp (i\sigma \phi _j),$ $A_{ij}^h\rightarrow
A_{ij}^h+i(\phi _i-\phi _j).$ So this model has a \textrm{U(1)}$\times $%
\textrm{U(1)} gauge symmetry. The nontriviality of $H_{string}$ arises from
the `dual' gauge fields, $A_{ij}^s$ and $A_{ij}^h,$ which satisfy
topological conditions: $\sum_cA_{ij}^s=\pm \pi \sum_{l\in c}(n_{l\uparrow
}^s-n_{l\downarrow }^s)$ and $\sum_cA_{ij}^h=\pm \pi \sum_{l\in c}n_l^h$ for
a closed loop $c$. 

\begin{figure}[t]
\centerline{\epsfxsize=3.4in\epsfbox{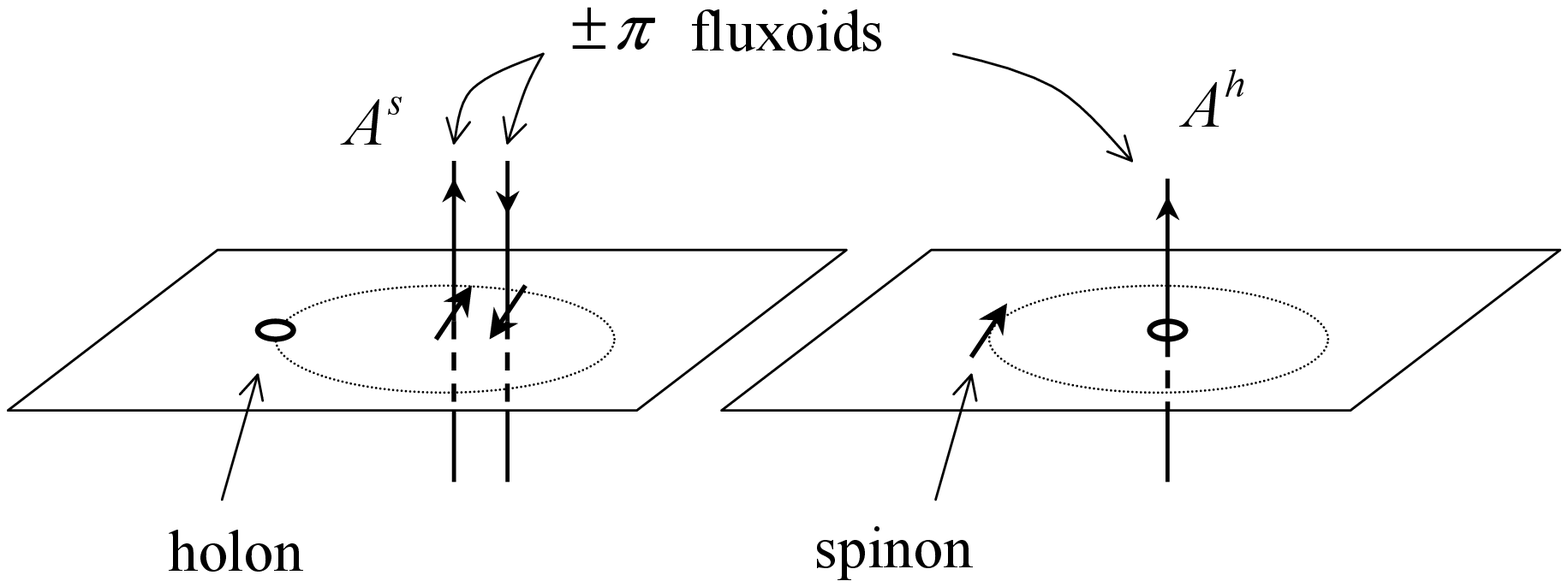}}
\caption{Dual topological gauge structure represented by $A^s_{ij}$ and $A^h_{ij}$
in the effective Hamiltonians, $H_h$ and $H_s$.}
\label{inter}
\end{figure}

The holons in (\ref{hh}) feel the presence of the spinons as quantized $\pi $
fluxoids through $A_{ij}^s$, which reflects the nonlocal frustrations of the
spin RVB background on the kinetic energy of the charge degrees of freedom,
due to the phase string effect. \emph{Vice versa }the spinons also perceive
the doped holes as $\pi $ flux quanta through $A_{ij}^h,$ which represents
the dynamic frustrations of the doped holes on the spin degrees of freedom.
Such a dual and mutual gauge structure is schematically illustrated in Fig.
2. In the following, the phase diagram determined by such a topological
gauge structure will be present, which unifies the AF states, the
superconducting phase, and the so-called pseudogap phase within a single
framework, as shown in Fig. 3.
\section{Phase diagram}

\subsection{Half-filling and dilute doping}

\emph{Half-filling. }At half-filling, the holon number is zero and $%
H_{string}$ reduces to $H_s.$ With $A_{ij}^h=0,$ $H_s$ here is equivalent to
the Schwinger-boson mean-field Hamiltonian \cite{aa}, which describes AF
correlations of the cuprates fairly well over a wide temperature range $\sim
1,000K.$ As a matter of fact, using the mean-field wavefunction of $H_s,$ we
can numerically calculate the ground-state energy and the AF magnetization
by the variational Monte Carlo method, which agree with the exact
diagonalization results of the Heisenberg model extremely well. All of these
indicate that both long-range and short-range spin correlations are well
captured by the bosonic RVB theory at half-filling.

\emph{Dilute doping. }Now consider hole doping into such an AF spin
background. As pointed out in the Introduction, the motion of holes will
generally induce the phase string effect. Since here we focus on the dilute
limit of the hole concentration, the phase string effect will mainly
influence the hole dynamics, without drastically affecting the spin part.
The holes will then be self-trapped or localized by the phase string effect
in such a dilute doping regime \cite{1-hole,string3,string4}, as to be discussed 
below. Such a localization is purely a consequence of the Mott
insulators upon doping, unlike the conventional Anderson localization in the
presence of disorders.

\emph{Localization. }The localization of holes can be mathematically
determined by the topological gauge structure of the phase string model.
Note that the long range AF order is realized by the spinon Bose
condensation, i.e., $\left\langle b_{i\sigma }\right\rangle \neq 0,$ at
half-filling. If the spinon condensation persists into a dilute doping
regime, each holon will induce a vortex in the spinon condensate through $%
A_{ij}^h$ in (\ref{hs}), which costs a logarithmically divergent energy. As
the result, each holon will have to induce an antivortex from the spinon
condensate and be confined to the latter to form a hole-dipole object in the
spin ordered (spinon condensed) phase \cite{string3}. In this hole-dipole entity, the holon
sits at one of two \emph{poles} instead of the \emph{center} of a dipole. It
can be shown that the effective mass of the induced antivortex is infinity
such that the whole hole-dipole object is self-trapped in space. This hole
localized phase is called \emph{holon confined phase }\cite{string3}, in
which no free holons will appear in the finite-energy spectrum.

\emph{Competing orders. }In such a localization regime, the kinetic energy
of holes is suppressed. Such a suppression of the kinetic energy is one of
the most essential features of the holon-confined phase. Without the balance
from the kinetic energy, the low-energy physics in this regime will be
determined by the potential energies. The latter will then decide various
competing orders at such a low-doping insulating phase.

First of all, at sufficiently low doping, the long-range AF order should
persist if a weak interlayer coupling is considered. But the freedom in the
directions of the hole-dipole moment will lead to the reduction of the N\'{e}%
el temperature $T_N$ as shown in Fig. 3. Based on the hole-dipole picture
and the renormalization group (RG) calculation we have determined \cite
{string3} the critical doping $x_0\simeq 0.03,$ at which the AF order
disappears at $T=0.$

Beyond $x_0$ or $T_N,$ the system is in a cluster spin glass phase with the
dipole moments being quenched randomly in space \cite{string3}. This phase
has been also observed in experiment. In obtaining such a phase, we have
assumed that hole-dipoles are self-trapped in space \emph{uniformly}. This
latter condition may be realized in real systems by impurities or disorders.

However, if there is no impurities or disorders, the uniform distribution of
the self-trapped hole-dipoles are usually not stable against the formation
of the stripes \cite{string4}. This is because the long-range dipole-dipole
interaction will dominate the low-energy phase as there is no competition
from the kinetic energy. Such a dipole-dipole interaction will lead to the
attraction among hole-dipoles with a certain alignment of their dipole
moments. Their subsequent collapse together with the ordered dipole moments
will lead to various stripe formations. Therefore, in the phase string
theory, inhomogeneous distribution of the charges is energetically very
competitive at low doping.

\begin{figure}[t]
\centerline{\epsfxsize=3.9in\epsfbox{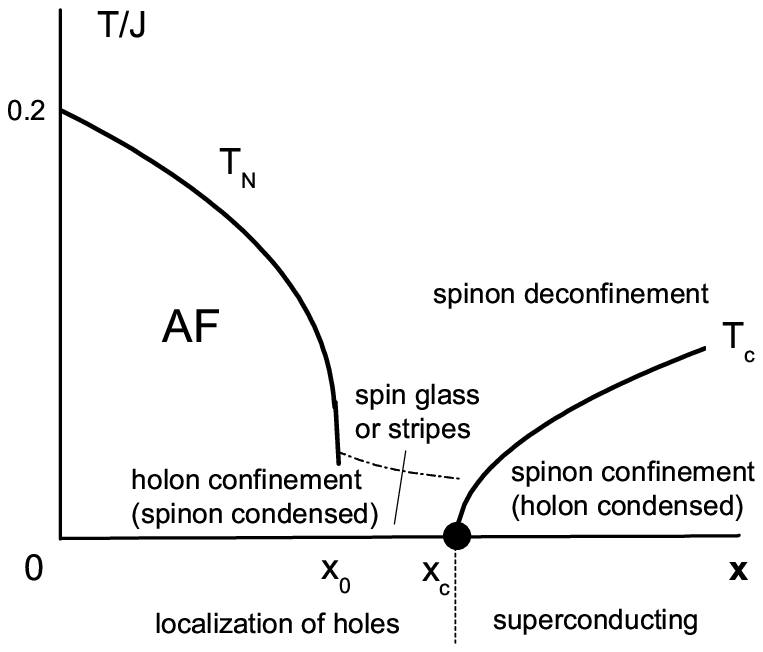}}
\caption{The phase diagram of the phase string model at small doping. The dual topological gauge
structure decides various phases through confinement-deconfinement procedures. The
RG calculations give rise to $x_0\sim 0.03$ and $x_c\sim 0.043$ (see text).}
\label{inter}
\end{figure}

\emph{Holon deconfinement.} With the further increase of doping, the sizes
of hole-dipoles will get larger and larger, and eventually a deconfinement
can occur at a critical doping $x_c$ \cite{string3}, beyond which single
holons will be unbound from their anti-vortex partners. This procedure is
like a KT transition in the $x$ axis and by the RG method we have determined 
$x_c\simeq 0.043$ at $T=0.$ In the following section, we will point out that
the ground state in the holon deconfined phase is a superconducting state.

\subsection{ Superconducting state}

We now consider the regime beyond the critical doping $x_c$. The deconfined
bosonic holons will experience a Bose condensation at low temperature in $%
H_h $. The corresponding ground state is a d-wave superconducting state as
to be discussed below.

\emph{d-wave pairing order parameter. }In the phase-string theory,
superconducting order parameter can be expressed as \cite{string5} 
\begin{equation}
\Delta _{ij}^{\mathrm{SC}}=\Delta _{ij}^0\langle e^{i\frac 12\left( \Phi
_i^s+\Phi _j^s\right) }\rangle ~,  \label{odlro}
\end{equation}
where $ij$ refer to nearest neighbor sites and the pairing amplitude 
\[
\Delta _{ij}^0\varpropto \Delta ^s\left\langle h_i^{\dagger }\right\rangle
\left\langle h_j^{\dagger }\right\rangle 
\]
in which the bosonic RVB pairing $\Delta ^s\neq 0$ and the holon
condensation $\left\langle h_i^{\dagger }\right\rangle \neq 0$ usually
occurs at relatively higher temperatures, $T_0$ and $T_v$, respectively. The
true superconducting transition will happen when $\langle e^{i\frac 12\left(
\Phi _i^s+\Phi _j^s\right) }\rangle \neq 0$, namely, when the phase
coherence is established. The phase factor $e^{i\frac 12\left( \Phi
_i^s+\Phi _j^s\right) }$also determines the d-wave symmetry of the pairing 
\cite{string5}.

\emph{Spinon as a vortex. }The phase $\Phi _i^s$ represents vortices bound
to spinons \cite{string6}. In the ground state, where spinons are all paired
up, $\Phi _i^s$ is canceled out and the phase coherence is realized in (\ref
{odlro}). Each unpaired spinon, in an excited state, will induce a vortex in
the order parameter through $\Phi _i^s$. Such an effect appears in $H_h$
through $A_{ij}^s$, which will cost a logarithmically divergent energy.

\emph{Confinement and elementary excitations. }Thus a single spinon as a
vortex cannot appear in the superconducting bulk alone. The excited spinons
must appear in vortex-antivortex pairs and be confined to form $S=1$ triplet
excitations \cite{string5,string6}. In contrast to the holon confinement in
the spin ordered phase with spinon condensation, the present spinon
confinement occurs in the holon condensed phase, due to the gauge field $%
A_{ij}^s$ in $H_h$.

Besides the spin triplet excitations composed of confined spinons, there is
another kind of elementary excitations, namely, a nodal quasiparticle
excitation. Such an excitation can be created by the electron operator $%
c_{i\sigma }$ as a composite of confined holon and spinon with a
phase-string phase factor \cite{string5}. Such an excitation is similar to
that in a BCS theory.

But there is a clear non-BCS feature here. Namely, the $S=1$ spin excitation
composed of two confined spinons, which will form a resonance-like peak at $%
E_g\sim xJ$ in the phase-string theory$,$ is \emph{independent }of the
quasiparticle excitations at the zeroth-order approximation.

\emph{Superconducting transition. }The deconfinement of spinon-vortex pairs
occurs at $T=T_c$, which is responsible for destroying the phase coherence
of the order parameter (\ref{odlro}) as free unbound spinon-vortices emerge.
In this way, the superconducting transition is naturally related to both the
spin dynamics and the phase coherence within a single unified framework. $T_c
$ determined in this description is found \cite{string6,shaw} to be scaled
with the spin characteristic energy scale $E_g$, in good agreement with the
cuprate superconductors.

\emph{Flux quantization at }$hc/2e$ \emph{and vortex core. }In the fermionic
RVB mean-field theory, the superconducting state corresponds to the holon
Bose condensation, where the magnetic flux quantization is usually at $hc/e$%
, as holons carry a charge +e. How to get the correct flux quantization at $%
hc/2e$ has been a challenging issue in a spin-charge separation framework
based on the fermionic RVB theory.

In the phase-string theory, although the framework is still based on a
spin-charge separation description, the minimal magnetic flux is found to be
quantized at $hc/2e,$ similar to the BCS theory. However, as a unique
prediction of the theory\cite{string6}, the flux quantization leads to the
trapping of a spinon inside the magnetic vortex core. We expect that the
free $S=1/2$ moment trapped inside a vortex core can be observed in the NMR
and other experimental measurements.

Furthermore, the vortex core is different from the one in the BCS theory.
Here the pairing amplitude $\Delta _{ij}^0$ can remain finite throughout the
vortex core, while the superconducting order parameter $\Delta _{ij}^{%
\mathrm{SC}}$ vanishes as the phase coherence is destroyed inside the core
within the a characteristic length scale. Such a `normal core state'
resembles the `pseudogap' phase above $T_c,$ to be discussed below.

\subsection{Pseudogap phase}

\emph{Mechanism: kinetic energy driven}. The spin superexchange energy is
favored when spins form bosonic RVB pairs at $T_0$ ($\sim J_s/k_B\sim 1,000K$
at $x\sim 0$). With the decrease of temperature, bosonic holons will start
to gain kinetic energy and eventually experience a Bose condensation at $T_v$
(in the Kosterlitz-Thouless sense). Such a $T_v$ will represent an onset
temperature for the pseudogap phase in the bosonic RVB theory, rather than
the superconducting transition temperature in the conventional fermionic RVB
theory. The pseudogap temperature in the latter theory is defined as the
onset temperature of the RVB pairing, and thus the pseudogap phase is
superexchange-energy-driven there, instead of kinetic energy driven in the
present description.

\emph{Amplitude forming of the superconducting order parameter. }As
discussed in the above section, the amplitude of the superconducting order
parameter $\Delta _0$ forms at $T=T_v$, whereas the phase coherence of the
order parameter can still remain absent until $T=T_c<T_v.$

In the pseudo-gap phase of $T_c<T<T_v,$ spin RVB pairs are ``charged'' and
become Cooper pairs, as holons are condensed. In this sense, one may say
that preformed Cooper pairs emerge in the pseudogap phase \cite{string7}.

\emph{Spinon vortices. }A unique prediction of the phase string theory in
the pseudo-gap phase is the presence of topological vortices which are
spinons. This may be easily seen based on (\ref{hh}). According to $H_h,$
when holons are condensed, each excited (unpaired) spinon will induce a
current vortex through the gauge field $A_{ij}^s,$ which also shows up in
the pairing order parameter as discussed before. These spinon vortices are
free and the pseudo-gap phase has been thus called the spontaneous vortex
phase \cite{string7}.

Experimentally the presence of free vortices will contribute to the Nernst
effect. Indeed, nontrivial Nernst effect has been observed \cite{nernst} in different
cuprate compounds in the pseudogap regime. It is very hard to interpret the
experimental results as due to the conventional superconducting fluctuations
as the observed Nernst signal can persist over a temperature range ten times
higher than $T_c$ in the underdoped regime. The spinon-vortices in the
present theory provides a natural explanation for the experiment.

\emph{Spin pseudogap.} Since the spin degrees of freedom is influenced by
the charge dynamics through the gauge field $A_{ij}^h$, the holon
condensation below $T_v$ will directly cause the qualitative change of spin
dynamics. Namely, it will lead to the suppression of the spin fluctuations
at the low-energy end, resembling a spin pseudogap feature, with the
spectral weight being pushed upwards to form a resonance-like peak structure
around a characteristic energy scale $E_g\sim \delta J.$ As discussed
before, this feature persists into the superconducting phase.

A pseudogap means that it is more difficult to excite spinons from the RVB
pair condensate. Since $A_{ij}^s$ is canceled out in the RVB state, in the
pseudogap phase the fluctuations of $A_{ij}^s$ are also reduced at low
energy. This in turn favors the holon condensation. Therefore,
self-consistently, the holon condensation and spin pseudogap are mutually
enforced by each other.

\section{Summary}

In this talk, I have discussed several fundamental issues of the doped Mott
insulator, including spin-charge separation, the gauge theory description of
couplings between two degrees of freedom, and short-range quantum
correlations which provide the driving force for the low-energy physics.

I have pointed out that the bosonic RVB pairing characterizes the crucial
short-range correlations at both half-filling and small doping. Once such a
short-range, high-energy correlations are determined, the low-energy,
long-range theory is fixed, which is known as the phase string theory, as
the motion of holes in such a bosonic RVB background always generates
irreparable phase string effect.

I have reviewed some very basic features obtained within the phase string
framework. The AF long-range ordered phase persists into a finite doping
below $x_0\sim 0.03,$ and then becomes cluster spin glass phase at $x>x_0$. In both
phases holes are localized or self-trapped by the phase string effect. We
found that the stripe state is also very competitive in this regime. Beyond
a critical doping at $x_c\sim 0.043$, the holes will become delocalized, and
the system experiences a superconducting transition at low temperature. In
such a d-wave superconducting state, spinons are confined to form $S=1$
triplet excitations. Spinon and holon are confined to form nodal
quasiparticles. $T_c$ is related to the phase coherence transition and is
scaled with the spin `resonance' energy $E_g\sim xJ.$ Above $T_c$, a
pseudogap phase is found in which free spinons behave like unbound vortices.
Several novel properties are revealed in both superconducting and pseudogap
phases, including the prediction that a free moment trapped inside a $hc/2e$
magnetic vortex core and the anomalous Nernst effect. All of these
interesting features and competing phases at different doping and
temperature are determined self-consistently by the unique dual topological
gauge structure in the phase string theory.    

\section*{Acknowledgments}

The recent part of this work has been done in collaboration with V.N.
Muthukumar, S.P. Kou, M. Shaw, and Y. Zhou. The support from NSFC grants is
also acknowledged.

\end{document}